\documentstyle[12pt]{article}

\newcommand{\resection}[1]{\setcounter{equation}{0}\section{#1}}
\newcommand{\EQ}{\begin{equation}}
\newcommand{\EN}{\end{equation}}
\newcommand{\EQN}{\[}
\newcommand{\ENN}{\]}
\newcommand{\EQA}{\begin{eqnarray}}
\newcommand{\ENA}{\end{eqnarray}}

\newcommand{\hs}{\hspace{0.1cm}}

\newcommand{\vtp}{\vartheta(p)}

\newcommand{\s}{\sigma}
\newcommand{\sm}{\sigma^{(m)}}

\newcommand{\eps}{\varepsilon}

\newcommand{\ra}{\rangle}

\newcommand{\goto}{\rightarrow}

\begin{document}
\oddsidemargin 5mm
\setcounter{page}{0}
\renewcommand{\thefootnote}{\fnsymbol{footnote}}
\newpage
\setcounter{page}{0}
\begin{titlepage}
\begin{flushright}
{\small E}N{\large S}{\Large L}{\large A}P{\small P}-L-381/92
\end{flushright}
\begin{flushright}
ITP - SB - 92 - 21
\end{flushright}
\vspace{0.5cm}
\begin{center}
{\large {\bf Determinant Representation for the Time Dependent Correlation
Functions in the XX0 Heisenberg Chain}} \\
\vspace{1cm}
{\bf F. Colomo$^{1}$,
A.G. Izergin$^{2}$
\footnote{Permanent Address: {\em Sankt-Petersburg Branch (POMI) of V.A.
Steklov Mathematical Institute of Russian Academy of Sciences, Fontanka
27, 191011 St.-Petersburg, Russia.}},
V.E. Korepin$^{3}$,
V. Tognetti$^{4}$ \\
\vspace{0.8cm}
$^1${\em I.N.F.N., Sezione di Firenze,
Largo E. Fermi 2, 50125 Firenze, Italy}\\
$^2${\em Forum: Project on Condensed Matter Theory of INFM, Firenze-Pisa,
Italy}\\
and\\
{\em Laboratoire de Physique Th\'eorique ENSLAPP, Ecole Normale
Superieure de Lyon, France}\\
$^3${\em  Institute for Theoretical Physics, State University of
New York at Stony Brook, NY-11794-3840, USA}\\
$^4${\em Dipartimento di Fisica, Universit\`a di Firenze, Largo E. Fermi
2, 50125 Firenze, Italy}}
\end{center}
\vspace{6mm}
\begin{abstract}
Time dependent correlation functions in the Heisenberg XX0 chain in
the external transverse magnetic field are calculated. For a finite chain
normalized mean values of local spin products are represented as
determinants
of $N\times N$ matrices, $N$ being the number of quasiparticles in the
corresponding eigenstate of the Hamiltonian. In the thermodynamical limit
(infinitely long chain), correlation functions are expressed in terms of
Fredholm determinants of linear integral operators.
\end{abstract}
\vspace{3mm}
\centerline{April 1992}

\bigskip

\centerline{REVISED VERSION}
\end{titlepage}

\newpage
\renewcommand{\thefootnote}{\arabic{footnote}}
\setcounter{footnote}{0}

\resection{Introduction}

XY model was introduced by E. Lieb,T. Schultz and D. Mattis \cite
{one}.Correlation
functions were studied in the model in great detail \cite{one}
- \cite{sevenprime}, but the problem of evaluation of large distance
asymptotics of temperature correlations
is still an open question. This problem was solved for impenetrable Bose
gas in
\cite{IIKnew}
- \cite{book} . In these papers it was shown that quantum
correlation
function is the $\tau$ function of classical completetly integrable
differential equation.
The sequence of calculations  in these papers is the following:
First correlation function is reprsented in the form of determinant of
Fredholm
integral operator. This permits to reduce the problem of evaluation of
asymptotics
 to Riemann-Hilbert problem. In the present paper we do only first step fo
isotropic XY
model.We represent temperature correaltion function as a determinant of
integral operator. In the next publication we shall formulate corresponding
Riemann-Hilbert problem and evaluate asymptotics of temperature
correlations.
Actually we should mention that asymptotics of of one paricular case of
 temperature correation was calculated in isotropic XY model.
Autocorrelation was
 represented as Fredholm determinant in \cite{sevenprime} , Riemann-Hilbert
problem
 was formulated in \cite{novok} . It was solved and asymptotics was
evaluated in
\cite{deift}. The nature of Fredhom determinant obtained in this paper
is different from \cite{sevenprime}

\bigskip

The XX0 chain is the isotropic case of the XY model \cite{one}, being also
the
free fermions point for the XXZ chain. The Hamiltonian describing the
nearest
neighbour interaction of local spins $\frac{1}{2}$  situated at the sites
of
the one-dimensional lattice in a constant transverse magnetic field $h$ is
\EQ
H(h)=-\sum_{m=1}^M\left[\s_x^{(m)}\s_x^{(m+1)}+\s_y^{(m)}\s_y^{(m+1)}+h\sm_z
\right]\hs.\label{ham}
\EN
The total number $M$ of sites is supposed to be even and periodical
boundary
conditions, $\s^{(M+1)}_s=\s_s^{(1)}$ ($s=x, y, z$), are imposed. Pauli
matrices are normalized as $\left(\sm_s\right)^2=1$.

The ferromagnetic state $\mid 0\ra\equiv\otimes_{m=1}^{M}\mid\uparrow\ra_m$
(all spins up) is an eigenstate of the Hamiltonian. All the other
$2^M-1$ eigenstates can be constructed by filling this ferromagnetic state
with $N$ quasiparticles ($N=1, 2, ... , M$) possessing quasimomenta $p_a$
($a=1, 2, ... , N$) and energies $\eps(p_a)$,
\EQ
\eps(p)\equiv\eps(p,h)=-4\cos p +2h\hs.\label{energy}
\EN
Due to periodical boundary conditions, one has the following condition
for the permitted values of quasimomenta
\EQ
e^{iMp_a}=(-1)^{N+1}\hs, \hspace{1.5cm} a=1, ... , N\hs.\label{bethe}
\EN
All the  momenta of the quasiparticles in a given eigenstate should be
different, so that, $e.g.$, for $N=M$, one gets in fact only one eigenstate
(which is just the other ferromagnetic state with all spins down,
$\mid 0'\ra=\otimes_{m=1}^M\mid \downarrow\ra_m$).

Due to the similarity transformation,
\EQA
&&H(h)\hs \goto\hs H(-h) = UH(h)U^{-1};\hspace{2.0cm} U=\prod_{m=1}^{M}
\sm_x\hs,\nonumber\\
&&\mid 0 \ra\hs\goto\hs\mid 0'\ra=U\mid 0 \ra\hs,\label{simil1}
\ENA
it is sufficient to consider only nonnegative magnetic fields, $h\geq 0$.
Furthermore the choice of  the minus sign at the r.h.s.
of eq. (\ref{ham}) is just a matter of convenience due to the property
\EQN
H(h)\hs \goto\hs -H(-h) = VH(h)V^{-1};\hspace{2.0cm} V=\prod_{m=1}^{M/2}
\s^{(2m)}_z\hs.\label{simil2}
\ENN

The model in the thermodynamic limit ($M$ $\goto$ $\infty$, $h$ fixed) is
the most interesting. For $h\geq h_c\equiv 2$, the ground state of the
Hamiltonian is just the ferromagnetic state $\mid 0 \rangle$.
For magnetic field smaller than the critical value, $0\leq h< h_c$,
the ground state $\mid\Omega\ra$ is obtained by filling the ferromagnetic
state with quasiparticles
possessing all the allowed values of momenta inside the Fermi zone,
$-k_F\leq p_a\leq k_F$, where the Fermi
momentum $k_F$ is defined by the requirement $\eps(k_F)=0$:
\EQ
k_F=\arccos\left(\frac{h}{2}\right)\hspace{1.5cm}0\leq
h<h_c\hs.\label{fermi}
\EN
In the thermodynamical limit ($M\goto \infty$) the number $N_0$ of
quasiparticles in the ground state is going to infinity, $N_0\goto \infty$,
``density'' $D\equiv N_0/M$ remaining fixed.
At zero magnetic field $k_F=\frac{\pi}{2}$, and there are $\frac{M}{2}$
quasiparticles in the ground state, magnetization being equal to zero.

At non zero temperatures $T>0$, the distribution of quasiparticles in the
momentum space is $\vtp/2\pi$ where $\vtp$ is the Fermi weight
\EQ
\vtp\equiv\vartheta(p,h,T)=\frac{1}{1+\exp\left[\frac{\eps(p)}{T}\right]}\hs